\documentclass[aps,prl,tightenlines,onecolumn,groupedaddress,superscriptaddress,showpacs]{revtex4}
\usepackage{amssymb}
\usepackage{color,graphicx}
\usepackage{lmodern}
\usepackage{amsmath}
\usepackage{amsbsy}
\usepackage{amsthm}
\usepackage{float}
\usepackage{bbm}
\usepackage{bm}
\usepackage{epsfig}
\usepackage{caption}
\usepackage{graphicx}
\usepackage{subfig}
\usepackage{lipsum}

\newcommand{\be}{\begin{equation}}
\newcommand{\ee}{\end{equation}}
\newcommand{\beq}{\begin{eqnarray}}
\newcommand{\eeq}{\end{eqnarray}}
\newcommand{\bra}[1]{\ensuremath{\langle #1 |}}
\newcommand{\ket}[1]{\ensuremath{| #1 \rangle}}

\begin{document}
\title{Leggett-Garg Inequality for a Two-Level System under Decoherence}

\author{Nasim Shahmansoori}
\email[]{shahmansoori@ch.sharif.edu}
\affiliation{Research Group on Foundations of Quantum Theory and Information,
Department of Chemistry, Sharif University of Technology
P.O.Box 11365-9516, Tehran, Iran}
\author{Afshin Shafiee}
\email[Corresponding Author:~]{shafiee@sharif.edu}
\affiliation{Research Group on Foundations of Quantum Theory and Information,
Department of Chemistry, Sharif University of Technology
P.O.Box 11365-9516, Tehran, Iran}

\affiliation{School of Physics, Institute for Research in Fundamental Sciences (IPM), P.O.Box 19395-5531 Tehran, Iran}

\begin{abstract}
We consider a macroscopic quantum system subjected to an asymmetric double-well potential in a harmonic environment. By using a time-dependent approach, we calculate tunneling probabilities for the system which contain oscillation effects. To show how one can decide between quantum mechanics and the implications of macrorealism assumptions, a given form of Leggett-Garg inequality is considered. The violation of this inequality occurs for a broader range of the system-environment interactions, compared to previous results obtained for two-level systems. Assuming that the coupling strength between the system and the environment can be controlled with time, one can see the violation even for strong decoherence effects. We also investigate the variation of the tilt/tunneling parameters on the violation of Leggett-Garg inequality.
\end{abstract}
\maketitle

\section{Introduction}
Extrapolating the laws of quantum mechanics ($QM$), up to the scale of everyday objects, means that objects composed of many atoms exist in quantum superpositions of macroscopically distinct states. In 1935, Schrodinger attempted to demonstrate the counter-intuitive implications of $QM$ using a thought experiment in which a cat is put in a quantum superposition of alive and dead states ~\cite{[1a]}. The idea remained theoretical until 1980s, when much progress has been made in demonstrating the macroscopic quantum behavior of various systems such as superconductors ~\cite{[2a],[4a],[5a],[6a]}, nanoscale magnets ~\cite{[7a],[8a]}, laser-cooled trapped ions ~\cite{[10a]}, photons in a microwave cavity ~\cite{[11a]} and macromolecules ~\cite{[12a]}. 

A typical double-well potential system provides a unique opportunity to study the fundamental behavior of a macroscopic quantum system (MQS), specially macroscopic quantum tunneling. In the context of a double-well potential, Schrodinger's cat describes a state in which macroscopic system (macrosystem) simultaneously occupies both wells. There are also studies focused on decoherence effects in double-well potentials. Huang \textit{et al.} showed that decoherence due to the interactions of atoms with the electromagnetic vacuum can cause the defeating of Schrodinger cat-like states~\cite{[6]}. Thermal effects ~\cite{[7]} and dissipation ~\cite{[8]} constitute some sources of decoherence and can suppress tunneling between wells ~\cite{[9],[10]}. In addition, double-well potentials have been extensively applied in many branches of physics. For example, it appears in the dynamics of Bose-Einstein condensates, the recent developments of ion trap technology, the ultracold trapped atoms theory and its applications ~\cite{[1],[2],[3],[4],[5]}.

Such a situation brings in mind the question of how the everyday macroscopic world works. The Leggett-Garg inequality (LGI) provides a criterion to investigate the existence of macroscopic coherence and thereby test the applicability of $QM$ as we scale from the micro- to the macro-world ~\cite{[13a],[11]}. LGIs are based on two assumptions, macroscopic definiteness and noninvasive measurability. Violation of LGI implies either the absence of a realistic description of the system or the impossibility of measuring the system without disturbing it. $QM$ violates different forms of LGIs. A number of experimental tests and violations of these inequalities have been demonstrated in recent years ~\cite{[12],[13]}. Leggett and Garg initially proposed an rf-SQUID flux qubit as a promising system to test their inequalities ~\cite{[11]}, which was later improved by Tesche ~\cite{[14]}. The first measured violation of a type of LGI was reported by Palacious-Laloy \textit{et al.} ~\cite{[15]}. Palacios-Laloy \textit{et al.} found that LGI is violated by their qubit with the conclusion that their system could not admit a realistic, non-invasively-measurable description. Recently, several experimental tests of LGIs were implemented, all of which confirm the predicted violations in accordance with the fundamental laws of $QM$ ~\cite{[15],[2b],[3b],[4b],[6b],[7b],[8b],[9b]}. Most of these experiments were weak measurements, where the effects of the measured back-action in a sequential set up are minimized ~\cite{[10b]}.  

In this article we examine LGI regarding an asymmetric double-well potential in a harmonic environment. To do this, we consider the effect of the environment as a perturbation on the system. For symmetric double-well potentials considered as a two-level quantum system, it has been shown that $QM$ violates different forms of LGI ~\cite{[16]}. Moreover, no violation occurs, when strong decoherence is at work. According to our calculations for an asymmetric double-well potential, it is possible to see the violation, even for significant effects of the decoherence. This can be achieved by controlling the strength of interaction between the system and it's environment in different time domains. We also study the effects of the tilt/tunneling parameters on the violation of LGI. 

The structure of our paper is as follows. In section 2, we focus on an asymmetric double-well model. We introduce its Hamiltonian and consider the effects of the environment on it. Then, we calculate the tunneling probabilities to obtain time correlations. In section 3, the violation of a given LGI under decoherence is assessed. In section 4, we expound the violation range depending on different factors. Finally, in section 5, we conclude our results.   
\section{Asymmetric Double-Well Potential}

We consider a typical asymmetric double-well potential (FIG.1), where its asymmetric form is characterized by the parameter $\delta$. Here $\vert L\rangle$ ($\vert R\rangle$) denotes the state in which the macrosystem is localized in the left (right) well. We can control the macroscopic feature of the system by using dimensionless equations. 
A particle of mass $M$ passes through a potential which has the characteristic length $R_{0}$ and the characteristic energy $U_{0}$, defined as the units of the length and the energy, respectively. The corresponding characteristic time can be defined as $T_{0} = R_{0}/(U_{0}/M)^{1/2}$ which we consider as the time needed for a particle of mass $M$ to pass the distance $R_{0}$ with kinetic energy of the order of $U_{0}$. Likewise, the unit of the momentum is taken as $P_{0} = (MU_{0})^{1/2}$. We then define the dynamical variables, $q$, $p$ and $t$,
as $R/R_{0}$, $P/P_{0}$ and $T/T_{0}$ respectively. Instead of the plank's constant, a new dimensionless parameter $\tilde{h}$, is defined based on the commutation relation of $p$ and $q$ in units of action $U_{0}\tau_{0}$:
 \be
 \label{1}
 \tilde{h}=\dfrac{\hbar}{P_{0}R_{0}}. 
 \ee
 The value of the new parameter $\tilde{h}$ quantitatively characterizes the macroscopicity of the system. So that, for small values of $\tilde{h}$, the dynamic is more quasi-classical. Yet, to detect the quantum tunneling effect, $\tilde{h}$ should not be too small. Considering an asymmetric double-well potential, we assume that the value of $\tilde{h}$ is about 0.1 to support the macroscopic quantum trait of the system, in a quasi-classical situation. Here, we are only interested in the macroscopic quantum regime, where the system not only exhibits quantum oscillations (as for quantum and thus $\tilde{h}$￼ is not too small), but also involves a large number of dynamical degrees of freedom (as for macroscopic systems and thus $\tilde{h}$￼ is not too close to 1). The typical range of ￼ for a macroscopic quantum system is $0.01-0.1$. For most discussed macroscopic systems, i.e. SQUIDs and liquid He, the value of $\tilde{h}$ is estimated as 0.1 and 0.15, respectively. We chose the typical value of ￼ $\tilde{h}= 0.1$, so our approach does correspond to the macroscopic regime ~\cite{[16]}.
\begin{figure}[H]
\centering
\includegraphics[scale=0.7]{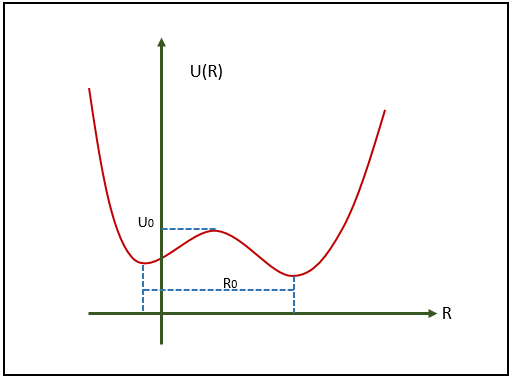}
\caption{Asymmetric double well potential} \label{fig1}
\end{figure}
 At enough low temperatures, energy states are confined in two-dimensional Hilbert space. When the macrosystem is isolated from its environment, it can be described effectively by the following Hamiltonian:
\begin{equation}
\label{2}
H=\frac{\tilde{h}}{2}
\begin{pmatrix}
\delta&\Delta\\\Delta&-\delta
\end{pmatrix}
\end{equation}
where $\delta=(E_{L}-E_{R})/\tilde{h}$ is a measure of the tilt, and $\Delta$ is a measure of the strength of the tunneling between the two wells. The eigenvalues of the Hamiltonian (2) are $\Omega_{mn}=\pm\frac{\tilde{h}}{2}(\delta^{2}+\Delta^{2})^{\frac{1}{2}} $ and the eigenstates of this Hamiltonian are:
\begin{subequations}
\begin{align}
\label{3}
\vert 0\rangle&=-\cos\theta\vert R\rangle+\sin\theta\vert L\rangle\\
\vert 1\rangle&=\sin\theta\vert R\rangle+\cos\theta\vert L\rangle ,
\end{align}
\end{subequations}
where $\theta=Arc\tan(\sqrt{((\delta^{2}+\Delta^{2})^{\frac{1}{2}}+\delta)/2(\delta^{2}+\Delta^{2})^{\frac{1}{2}}})$. Consequently, we have 
\begin{subequations}
\begin{align}
\label{4}
\vert R\rangle&=a\vert 0\rangle+b\vert 1\rangle\\
\vert L\rangle&=a^{\prime}\vert 0\rangle+b^{\prime}\vert 1\rangle ,
\end{align}
\end{subequations} 
where $a^{\prime}=b=\sin\theta$ and $b^{\prime}=-a=\cos\theta$. One can easily show that the probability of the tunneling from the left to the right well is
\be
\label{5}
P_{R}=\dfrac{\Delta^{2}}{\Delta^{2}+\delta^{2}}\sin^{2}[(\Delta^{2}+\delta^{2})^{\frac{1}{2}}\frac{t}{2}],
\ee
which is independent of $\tilde{h}$ and contains oscillation effects. Nevertheless, to deal with real systems, the inevitable effects of the environment should be considered. So, in order to retain oscillation effects and therefore the macroscopic quantum coherence, we consider the effects of the environment as a kind of perturbation on the system. We define $\ket{\alpha}$ and $\varepsilon_{\alpha}$ as the energy eigenstates and the energy eigenvalues of the environment, respectively. Apparently, the environment is assumed to be a bosonic field. The ground state of $\hat{H}_{\varepsilon}$, environment Hamiltonian is $\vert vac\rangle$ and $\vert\alpha\rangle=b_{\alpha}^{\dagger}\vert vac\rangle$ is the state with a single boson $\alpha$. The state $\ket{n,vac}\rangle$ is an eigenstate of $\hat{H_{0}}=\hat{H_{s}}+\hat{H_{\varepsilon}}$ with energy $E_{n}$. We define the interaction Hamiltonian, $\hat{H}_{s\varepsilon}$, as a perturbation on the system due to the environment
\be
\hat{H}_{s\varepsilon}=-\sum_{\alpha}(\omega_{\alpha}^{2}f_{\alpha}(\hat{q})\hat{x}_{\alpha}+\dfrac{1}{2}\omega_{\alpha}^{2}f_{\alpha}^{2}(\hat{q})),
\ee
where $f_{\alpha}(\hat{q})$ is an arbitrary function of $q$, depending on how the macrosystem exerts force on the environmental oscillators. We assume that the interaction model is bilinear i.e., $f_{\alpha}(q)=\gamma_{\alpha} q$, where $\gamma_{\alpha}$ is the coupling strength. Here, $\omega_{\alpha}$ is the frequency of the particle $\alpha$ in the environment. The time evolution of the entire system is studied by the perturbation theory. To do so, we are going to calculate the probability of finding the macrosystem in each well. This could be defined as

\be
\label{6}
P_{R}=\vert\langle R\vert\Psi(t)\rangle\rangle\vert^{2},
\ee
where $\ket{\Psi(t)}\rangle$, is the quantum state of the entire system at time $t$:
\be
\label{7}
\vert\Psi(t)\rangle\rangle=\sum_{i}\vert i\rangle\langle i\vert e^{-i\hat{H}_{\circ}t/\tilde{h}}\hat{U}_{I}\vert\Psi(0)\rangle.
\ee 
Here, $\hat{U_{I}}$ is the time-evolution operator in the interaction picture, given by $\hat{U}_{I}(t)=exp(-i\hat{H}_{0}t/\tilde{h})exp(-i\hat{H}t/\tilde{h})$ where $\hat{H}=\hat{H}_{0}+\hat{H}_{s\varepsilon}$. The relation (8) could be written in the following form:
\be
\label{8}
\vert\Psi(t)\rangle\rangle=\sum_{i}e^{-iE_{i}t/\tilde{h}}\vert i\rangle\vert\chi_{i}(t)\rangle ,
\ee
where $i=R, L$ and $\vert\chi_{i}(t)\rangle=\langle i\vert e^{-i\hat{H}_{\varepsilon}t/\tilde{h}}\hat{U}_{I}\vert\Psi(0)\rangle$. Hence, we have
\be
\label{9}
 P_{R}=\langle\chi_{R}(t)\vert\chi_{R}(t)\rangle .
\ee
The time evolution operator $\hat{U_{I}}$ could be expanded up to the second order with respect to the interaction Hamiltonian $\hat{H_{s\varepsilon}}$ as:
\be
\label{10}
\hat{U}_{I}(t)\simeq 1-\frac{i}{\tilde{h}}\int_{0}^{t}dt_{1}\hat{H}_{s\varepsilon}(t_{1})-\frac{1}{\tilde{h}^{2}}\int_{0}^{t}dt_{2}\int_{0}^{t_{2}}dt_{1}\hat{H}_{s\varepsilon}(t_{2})\hat{H}_{s\varepsilon}(t_{1}),
\ee
 where $\hat{H}_{s\varepsilon}(t)=e^{i\hat{H}_{0}t/\tilde{h}}\hat{H}_{s\varepsilon}e^{-i\hat{H}_{0}t/\tilde{h}}$. In (11), $\hat{U}_{I}(t)$ contains the following terms:
\beq
\label{11}
\hat{U}_{vac}(t)&=&-\frac{i}{\tilde{h}}\int_{0}^{t}\dfrac{1}{2}\sum_{\alpha}\omega_{\alpha}^{2}\hat{f}_{\alpha}^{2}(t_{1})dt_{1}-\frac{1}{2\tilde{h}}\sum_{\alpha}\int_{0}^{t}dt_{2}\int_{0}^{t_{2}}dt_{1}\hat{f}_{\alpha}(t_{2})e^{-i(t_{2}-t_{1})\omega_{\alpha}}\hat{f}_{\alpha}(t_{1}),\\
\hat{U}_{\alpha}(t)&=&\frac{i}{\sqrt{2}\tilde{h}}\int_{0}^{t}dt_{1}e^{-i\omega_{\alpha}t}\hat{f}_{\alpha}(t_{1}),\\
\hat{U}_{\alpha\beta}(t)&=&-\frac{1}{2\tilde{h}}\int_{0}^{t}dt_{2}\int_{0}^{t_{2}}dt_{1}\hat{f}_{\beta}(t_{2})e^{-i\omega_{\beta}t_{2}+i\omega_{\alpha}t_{1}}\hat{f}_{\alpha}(t_{1}).
\eeq 
The time-operator, $\hat{f}_{\alpha}(t)$ is defined in the interaction picture for $\hat{f}_{\alpha}(\hat{q})$. Using the relations (12)-(14) one can show that:
\begin{eqnarray}
\label{12}
\vert\chi_{R}(t)\rangle&=&\langle R\vert e^{-i\hat{H}_{\varepsilon}t/\tilde{h}}\hat{U}_{I}(t)\vert\Psi(0)\rangle\rangle \nonumber \\
&=&\vert vac\rangle\langle R\vert \hat{U}_{vac}\vert\psi_{s}(0)\rangle +\sum_{\alpha} e^{-i\omega_{\alpha}t}\vert\alpha\rangle\langle R\vert \hat{U}_{\alpha}\vert\psi_{s}(0)\rangle \nonumber \\
&&+\sum_{\alpha\beta} e^{-i(\omega_{\alpha}+\omega_{\beta})t}\vert\alpha\beta\rangle\langle R\vert \hat{U}_{\alpha\beta}(t)\vert\psi_{s}(0)\rangle .
\end{eqnarray}
If $\psi_{s}(0)=\vert L\rangle$, one gets;
\begin{eqnarray}
\label{13}
 P_{R}=\langle \chi_{R}\vert\chi_{R}\rangle=a^{2}b^{2}\vert\langle 0\vert \hat{U}_{vac}\vert 0\rangle\vert^{2} +{a^{\prime}}^{2}{b^\prime}^{2}\vert\langle 1\vert \hat{U}_{vac}\vert 1\rangle\vert^{2}\nonumber\\
+ 2aa^{\prime} bb^{\prime}\Re\langle 0\vert \hat{U}_{vac}\vert 0\rangle^{\ast}\langle 1\vert \hat{U}_{vac}\vert 1\rangle +{a^{\prime}}^2b^{2}\vert\langle 1\vert \hat{U}_{\alpha}\vert 0\rangle\vert^{2},
\end{eqnarray}
where $\Re$ denotes the real part. We have also used the relations (7a) and (7b) for the states $\ket{L}$ and $\ket{R}$.

In the tilted double-well potential calculations show that the elements $\bra{0}\hat{U}_{vac}\ket{1}$, $\bra{0}\hat{U}_{\alpha}\ket{0}$ and $\bra{1}\hat{U}_{\alpha}\ket{1}$ should be zero. The detailed results are given in appendix A.

Here, we used the following assumptions, appropriate in our case:\\ 
A1. The higher orders of $f_{\alpha}^{2}$ can be neglected, so \mbox{$\hat{U}_{\alpha\beta}=0$}.\\
A2. The frequency distribution $J(\omega)=\pi/2\sum_{\alpha}\omega_{\alpha}^{3}\gamma_{\alpha}^{2}\delta(\omega-\omega_{\alpha})$ of the environment is ohmic. This means that $J(\Omega_{mn})=\eta\Omega_{mn}$ where $\eta$ is a measure of the strength of the interaction between the macrosystem and the environment. We assume that $\eta$ is a small constant ($\eta\ll\tilde{h}$). \\
A3. The distribution $J(\Omega_{mn})$ is always positive. Thus $J(0)=0$ and $J(-\Omega_{mn})=0$ where $\Omega_{mn}=-\Omega_{nm}$.\\
 With all these assumptions in mind, if we suppose that the macrosystem is initially in the state $\vert L,vac\rangle\rangle$, the tunneling probability can be obtained as (see appendix B):
\be
\label{14}
P_{L\rightarrow R}=\sin^{2}\theta+(\sin^{2}\theta\cos2\theta)e^{-\Gamma_{1}t}-2\sin^{2}\theta\cos^{2}\theta\cos(\bar{\Omega}_{10}t)e^{-\Gamma_{1}t/2},
\ee
where $\bar{\Omega}_{10}=(\delta E_{1}-\delta E_{0})/\tilde{h}$ and $\Gamma_{1}$ is obtained according to fermi's golden rule.
\begin{eqnarray}
\label{15}
\Gamma_{n}&=&\frac{2}{\tilde{h}}\sum_{m}\vert f_{mn}\vert^{2}J(\Omega_{nm})\theta(\Omega_{nm)}
\end{eqnarray}
Here, $\Gamma_{1}^{-1}$ is the life time of the shifted energy $E_{1}+\delta E_{1}$. We also define $f_{mn}=\bra{m}f(\hat{q})\ket{n}$. This tunneling result shows that there is a decay factor $e^{-\Gamma_{1}t/2}$ that reduces the strength of the oscillation due to the decoherence (dephasing) effects. In order to diminish the effect of $e^{-\Gamma_{1}t/2}$, we consider the \textit{principal time} domain, which requires that $\Gamma_{1}t\ll1$. This assumption helps to conserve oscillation between the wells.\\
In the same way, one can calculate other probabilities. For example, when the macrosystem is in the state $\ket{R}$ initially, the probability that it could be found in the state $\ket{L}$ at time $t$ is denoted by $P_{R\rightarrow L}$. Taking into account the other probabilities $P_{R\rightarrow R}$ and $P_{L\rightarrow L}$, one can show that:
\begin{eqnarray}
\label{16}
P_{R\rightarrow L}&=&\cos^{2}\theta-\cos^{2}\theta\cos2\theta e^{-\Gamma_{1}t}-2\sin^{2}\theta\cos^{2}\theta \cos(\bar{\Omega}_{10}t) e^{-\Gamma_{1}t/2},\\
P_{R\rightarrow R}&=&\cos^{2}\theta-\sin^{2}\theta\cos2\theta e^{-\Gamma_{1}t}+2\sin^{2}\theta\cos^{2}\theta\cos(\bar{\Omega}_{10}t)e^{-\Gamma_{1}t/2}\\
P_{L\rightarrow L}&=&\sin^{2}\theta+(\cos^{2}\theta\cos2\theta)e^{-\Gamma_{1}t}+2\sin^{2}\theta\cos^{2}\theta\cos(\bar{\Omega}_{10}t)e^{-\Gamma_{1}t/2}.
\end{eqnarray}
\section{Violation of Leggett-Garg Inequality Under Decoherence} 
There are two main assumptions underlying any LG-type inequality, known collectively as macrorealism ($MR$) criteria. The assumption of $MR$ demands that, first, one can assign definite states to a macrosystem, so that it could be actually in one of these states independent of any observation. Second, it requires the non-invasive measurability of such macrostates which should not be affected, when they are measured. LGI serves to examine quantitatively whether the theories satisfying $MR$ are compatible with $QM$ or not. For this, we use the following LGI:
\begin{equation}
\label{17}
\mathcal{K}_{1}\equiv\vert C_{32}-C_{31}\vert+C_{21}\leq 1,
\end{equation}        
where the time-correlation function for the two-value variables $r$ and $q$ ($r, q=\pm1$) at three moments of time $t_{3}>t_{2}>t_{1}$ is defined as the following for the time sequences ${(i,j)=(3,2), (3,1), (2,1)}$:
\be
\label{18}
C_{ij}=\sum_{r,q=\pm1}rq P_{rt_{i},qt_{j}}.
\ee
For the symmetric double well potential and any other two-level system studied, these calculations show a maximum violation of $\mathcal{K}=3/2$, when the effect of decoherence is negligible \cite{[11]}. Now, let us assume that:
\be 
\label{19}
t_{3}-t_{2}=t_{2}-t_{1}=\frac{\tau}{\bar{\Omega}_{10}},\qquad\frac{\Gamma_{1}}{\bar{\Omega}_{10}}=\gamma,\qquad z=e^{-\gamma\tau}.
\ee
 Then, the estimation of a maximum value of $\gamma$ that violates LGI gives $\gamma=0.31$ \cite{[16]}. We also choose $\tau=\dfrac{\pi}{3}$, so that $\cos\bar{\Omega}_{10}(t_{2}-t_{1})=\frac{1}{2}$. Then, we have:
\begin{eqnarray}
\label{20}
\mathcal{K}_{1}&=&\vert P_{Rt_{3}\vert Rt_{2}}P_{Rt_{2}}+P_{Lt_{3}\vert Lt_{2}}P_{Lt_{2}}-P_{Rt_{3}\vert Lt_{2}}P_{Lt_{2}}\nonumber\\
&&-P_{Lt_{3}\vert Rt_{2}}P_{Rt_{2}}-(P_{Rt_{3}\vert Rt_{1}}P_{Rt_{1}}+P_{Lt_{3}\vert Lt_{1}}P_{Lt_{1}}\nonumber\\
&&-P_{Rt_{3}\vert Lt_{1}}P_{Lt_{1}}-P_{Lt_{3}\vert Rt_{1}}P_{Rt_{1}})\vert +P_{Rt_{2}\vert Rt_{1}}P_{Rt_{1}}\nonumber\\
&&+P_{Lt_{2}\vert Lt_{1}}P_{Lt_{1}}-P_{Rt_{2}\vert Lt_{1}}P_{Lt_{1}}-P_{Lt_{2}\vert Rt_{1}}P_{Rt_{1}},\qquad
\end{eqnarray}
where, \textit{e.g.}, $P_{Rt_{3}\vert Rt_{2}}=P_{Rt_{2}\rightarrow Rt_{3}}$ is the conditional probability that when the macrosystem is in the state $\ket{R}$ at $t_{2}$, it can be found in the same state $\ket{R}$ at $t_{3}$. Generally, we have $P_{rt_{i},qt_{j}}=P_{qt_{j}\vert rt_{i}}P_{rt_{i}}$ due to Bayesian rule where $P_{rt_{i}}$ is the single variable probability for the system being in the state $\ket{r}$ at $t_{i}(i=1,2,3)$. Conditional probabilities are given in relations (19) to (21), albeit without time labeling. Let us suppose that the macrosystem is initially in the state $\ket{L}$, so that $P_{Rt_{1}}=0$. Accordingly, $P_{Rt_{2}}$ is obtained from the following relation:
\begin{equation}
\label{21}
P_{Rt_{2}}=P_{Rt_{2}\vert Rt_{1}}P_{Rt_{1}}+P_{Rt_{2}\vert Lt_{1}}P_{Lt_{1}}=P_{Rt_{2}\vert Lt_{1}}.
\end{equation}
 Having into account the above considerations and using the relations (19) to (21), one can find that:  
\beq
\label{22}
\mathcal{K}_{1}&=&\vert(\sin^{2}\theta-\cos^{2}\theta+2\cos^{2}\theta\cos2\theta z_{a}+2\sin^{2}\theta\cos^{2}\theta z^{\frac{1}{2}}_{a})\nonumber\\
&&(\sin^{2}\theta+\cos^{2}\theta\cos2\theta z_{a}+\sin^{2}\theta\cos^{2}\theta z^{\frac{1}{2}}_{a})\nonumber\\
&&+(\cos^{2}\theta-\sin^{2}\theta-2\sin^{2}\theta\cos2\theta z_{a}+2\sin^{2}\theta\cos^{2}\theta z^{\frac{1}{2}}_{a})\nonumber\\
&&(\cos^{2}\theta-\cos^{2}\theta\cos2\theta z_{a}-\sin^{2}\theta\cos^{2}\theta z^{\frac{1}{2}}_{a})\nonumber\\
&&-(\sin^{2}\theta-\cos^{2}\theta+2\cos^{2}\theta\cos2\theta z^{2}_{b}-2\sin^{2}\theta\cos^{2}\theta z_{b})
\vert\nonumber\\
&&+\sin^{2}\theta-\cos^{2}\theta+2\cos^{2}\theta\cos2\theta z_{c}+2\sin^{2}\theta\cos^{2}\theta z^{\frac{1}{2}}_{c}.
\eeq
The time interaction factors $z_{a}$, $z_{b}$ and $z_{c}$ could be supposed to be equal or different, depending on the strength of the system-environment interaction in different time domains. Here, we first assume that they are all equal to each other, so that $z_{a}=z_{b}=z_{c}=z$. If we consider $\sin^{2}\theta=0.2$ and $\cos^{2}\theta=0.8$, at $z=1$ the inequality is violated, maximally. This situation is analogous to negligible decoherence. Yet, the important result is that for $0.5<z<1$, the inequality is violated too. This yields $0<\gamma<0.66$ which shows a broader range of violation compared to $\gamma=0.31$ for the symmetric double well potential and/or other proposed two-level systems \cite{[16],[17],[18]}. In FIG.2. $\mathcal{K}_{1}$ in (27) is plotted against $z$ for $\theta=26.6^{\circ}$. It is obvious that $\mathcal{K}_{1}$ increases as $z$ increases from 0 to 1. In FIG.3 $\mathcal{K}_{1}$ is plotted against $\sin^{2}\theta$ for $z=0.6$ (upper curve), $z=0.5$ (middle curve) and $z=0.4$ (lower curve). 

Of course, there are two other LGIs that our calculations show that they are not violated under the conditions considered above.
\beq
\label{17}
\mathcal{K}_{2}&\equiv &-( C_{32}+C_{21}+C_{31})\leq 1\\
-1\leq\mathcal{K}_{3}&\equiv & \dfrac{1}{2}( C_{43}+C_{32}+C_{21}-C_{41})\leq 1,
\eeq   
where $C_{ij}$s are defined according to relation (23). With the same way of calculating $\mathcal{K}_{1}$, we can calculate $\mathcal{K}_{2}$ and $\mathcal{K}_{3}$. The amounts of $\mathcal{K}_{2}$ and $\mathcal{K}_{3}$ versus $z$ are sketched in FIG. 4, for $\sin^{2}\theta=0.2$.

\begin{figure}[H]
\centering
\includegraphics[scale=0.7]{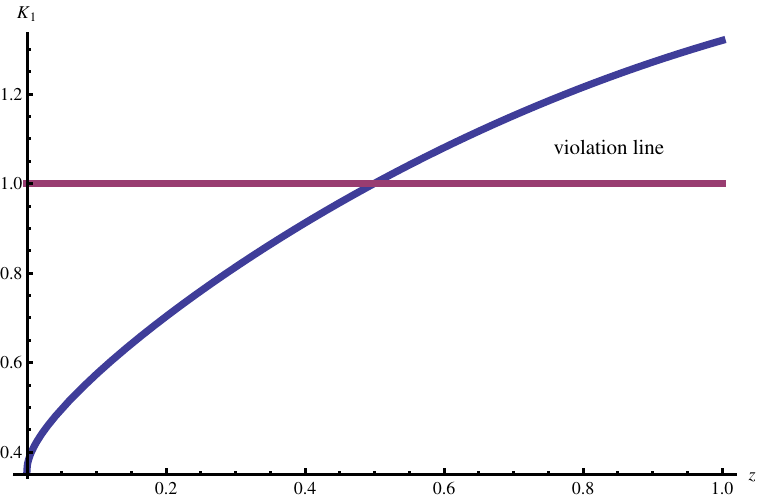}
\caption{The amount of $\mathcal{K}_{1}$ vs. $z$ for $\sin^{2}\theta=0.2$. By increasing $z$, $\mathcal{K}_{1}$ will increase for a constant amount of $\theta$. For all the $z$ upper than $0.5$ the inequality is violated.} \label{fig2}
\end{figure}
\begin{figure}[H]
\centering
\includegraphics[scale=0.7]{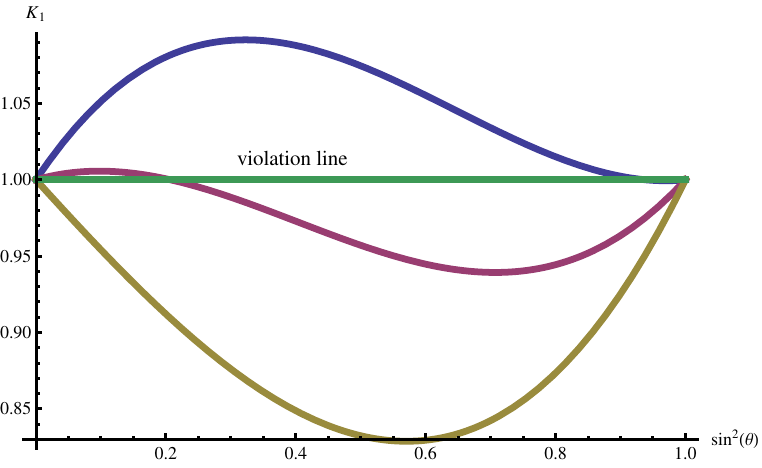}
\caption{The amount of $\mathcal{K}_{1}$ vs. $\sin^{2}\theta$ for three different values of $z=0.6$ (upper curve), $z=0.5$ (middle curve) and $z=0.4$ (lower curve).} \label{fig3}
\end{figure}
\begin{figure}[H]
\centering
\includegraphics[scale=0.5]{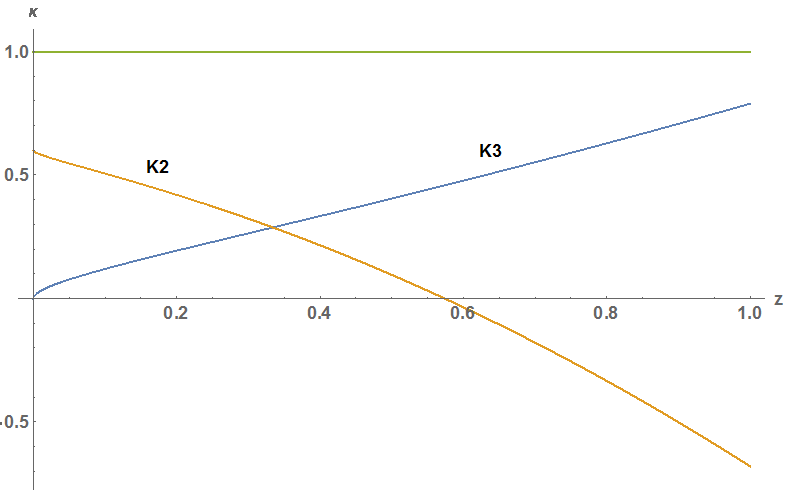}
\caption{The amounts of $\mathcal{K}_{2}$ and $\mathcal{K}_{3}$ vs. $z$ for $\sin^{2}\theta=0.2$.} \label{fig4}
\end{figure}

\section{Violation Probe With Varying Parameters}
We first examine the tilt/tunneling effect on the extent of violation. We define $d=\delta/\Delta$ where $\delta(\Delta)$ is the tilt (tunneling) parameter. The effect of $d$, depends on the strength of the interaction between the system and the environment. The strength can be controlled by the so-called parameter $\eta$. As is obvious in FIG. 5, when the strength of the interaction between the system and the environment is week (large values of $z$), the violation decreases by increasing $\delta/\Delta$ ratio, while the strong interaction between the system and the environment leads to an increase in violation. This is a guide line for experimental tests of LGI. Whether strong interactions prohibit LGI violation which can be interpreted as the compatibility of $MR$ and $QM$.

The tilt/tunneling effect on violation can be also assessed, using the functions $\sin\theta$ and $\cos\theta$ in (3a) and (3b). This can be seen in FIG. 6. Again, it is obvious that more violation are observed for large values of $z$ with weak interactions. 

As mentioned in the previous section $\tilde{h}$ is a parameter that controls macroscopicity in our calculations. here in FIG. 7 it is obvious that as the $\tilde{h}$ increases, the system is more quantum mechaniccal, the violation increases. This is in accordance with difficult observation of LGI violation for macroscopic systems. Whether for smaller values of $\tilde{h}$, more classical systems, the violation is also obseved.  

Now, we consider that the interaction strength could be controlled in different time domains of experiment. For example we assume that the system is isolated at the time domain $t_{2}-t_{1}$ (i.e., $z=z_{c}=1$ in (27)) and then it is allowed to interact with the environment. Then for all other times and for any other amounts of $z_{a}$ (defined for the time domain $t_{3}-t_{2}$) and $z_{b}$ (for $t_{3}-t_{1}$), the LGI is violated. This means that isolating the macrosystem in a given time domain causes the violation of LGI, even though the macrosystem is left open at all other times (see FIG. 4). This shows that the decoherence effect by itself has no role in diminishing the range of LGI violation. Yet, this is the time sequences of such effects which have the control role.

\section{Conclusion}
Considering a macrosystem prepared in a quasi-classical situation described by a tilted double-well potential, we studied the effect of the environment as a perturbation source. In this regime, the decoherence (dephasing) effects are reduced according to the so-called principal time domain in which $t\ll\Gamma_{1}^{-1}$. Calculations of the tunneling probabilities show that the coherency could be present, in spite of the interaction with the environment. To decide between the predictions of $QM$ and the requirements of $MR$, a type of LGI ($\mathcal{K}_{1}\leq1$) is considered in (22), when decoherence action is assumed to be present, but not so dominant. The violation of this inequality shows that the quantum behavior of a macrosystem could be present in more realistic situations. Even a small tilt in the double well potential can effect on the LGI violation. So, the key parameter $\gamma$ (characterizing the effect of dephasing) is improved from $\gamma=0.31$ in the previous works to $\gamma=0.66$.

Another important achievement is that by isolating the macrosystem at the time domain $t_{1}-t_{2}$ the LGI is violated even under strong decoherence effects at the other times. It should be mentioned that the time domain, $t_{1}-t_{2}$, can be considered as a very short time.These improvements are crucial for showing the violation of LGIs in the future proposed experiments. While the time domain $t_{1}-t_{3}$ can be very short. 

Also it is important to notice that, when the classical trait of the system is increased, which is illustrated by the larger values of $\gamma$, the assumption of non-invasive measurement is more possible to be violated. This means that time-correlations could be assumed to be achieved by higher time-ordered probabilities at the macro-level~\cite{[16]}. Due to the quantum calculations, this should be denied, since no three-variable joint probability could be defined for our model in quantum formalism from which one can obtain two-variable time-correlations. So, for broader ranges of violation due to large values of $\gamma$ $(\gamma\sim\Gamma_{1}\sim (\tilde{h} )^{-1})$ which shows the more classicality of the system, the violation of LGI features the violation of non-invasive measurability of the system in a more concrete way. It is legitimate to assume that physical properties of a macroscopic quantum system are definite and real. Yet, the violation of a typical LG inequality shows that any measurement on such a system should be invasive. Otherwise, the quantumness of the system could not be observed in such experiments~\cite{[19]}.  
\begin{figure}[H]
\centering
\includegraphics[scale=0.3]{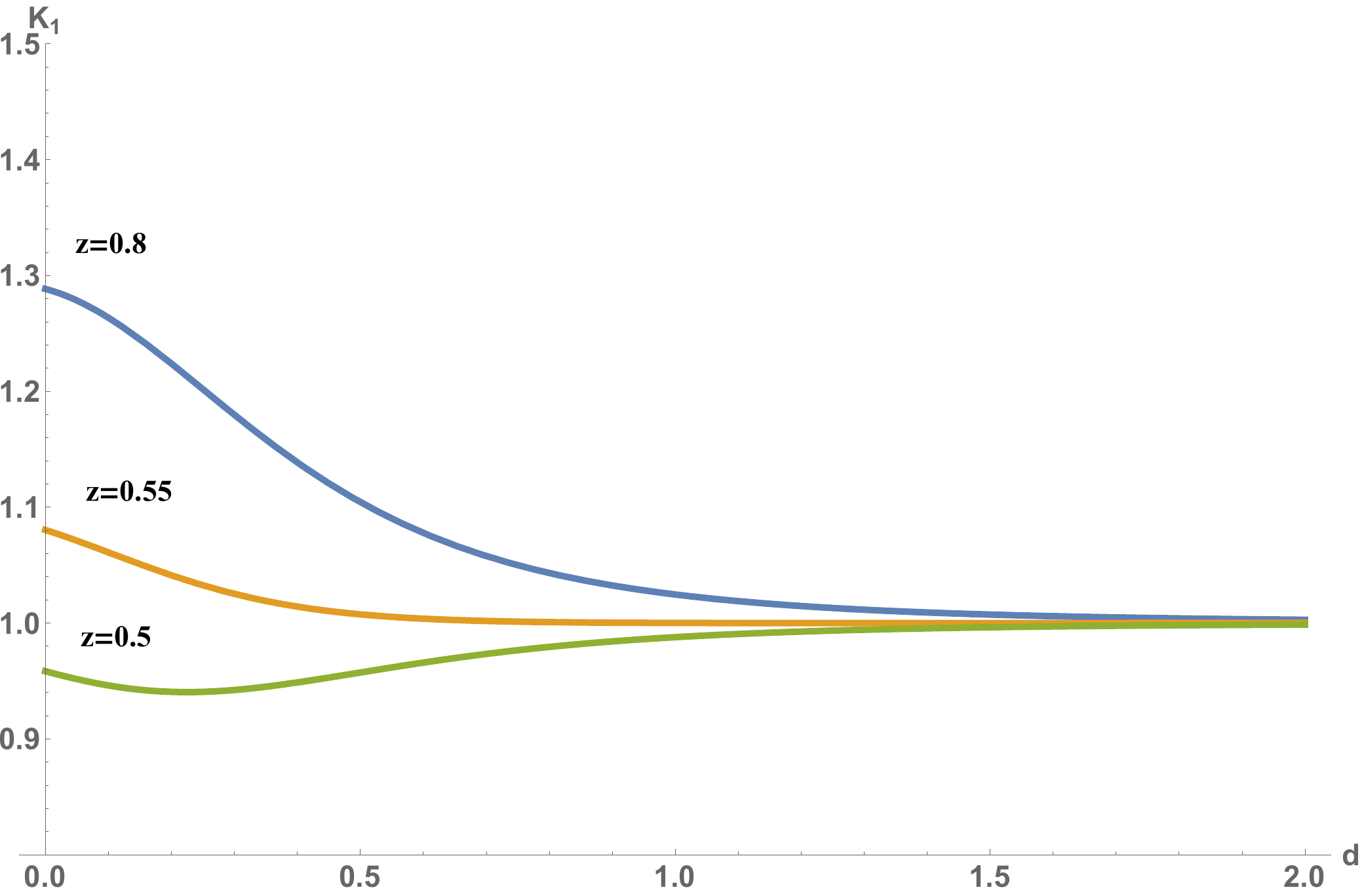}
\caption{The amount of $\mathcal{K}_{1}$ vs. $d=\delta/\Delta$ for $z=0.8$, upper curve, $z=0.55$, middle curve and $z=0.5$ lower curve} \label{fig3}
\end{figure}

\begin{figure}[H]
\centering
\includegraphics[scale=0.3]{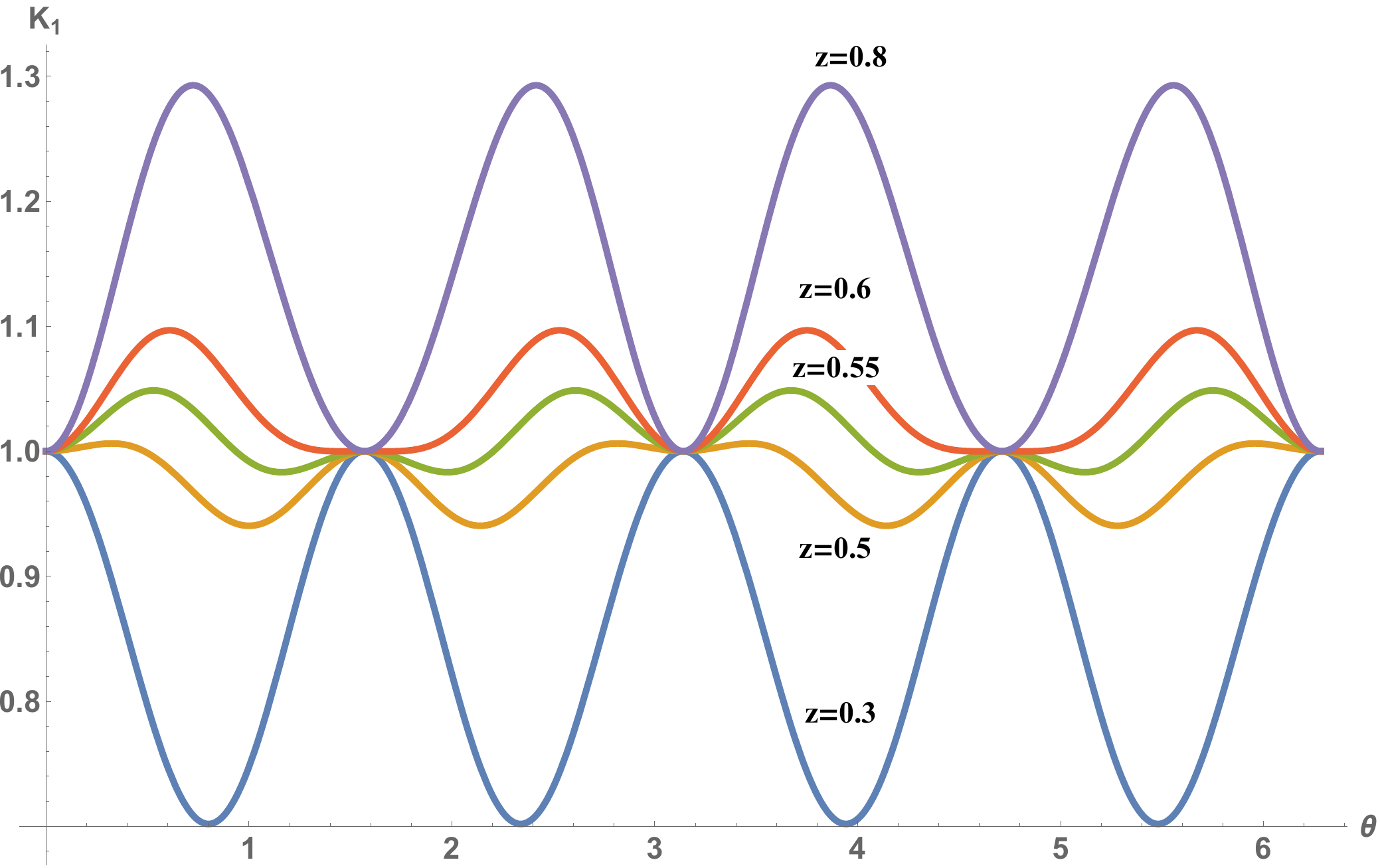}
\caption{The amount of $\mathcal{K}_{1}$ vs. $\theta$} \label{fig4}
\end{figure}

\begin{figure}[H]
\centering
\includegraphics[scale=0.3]{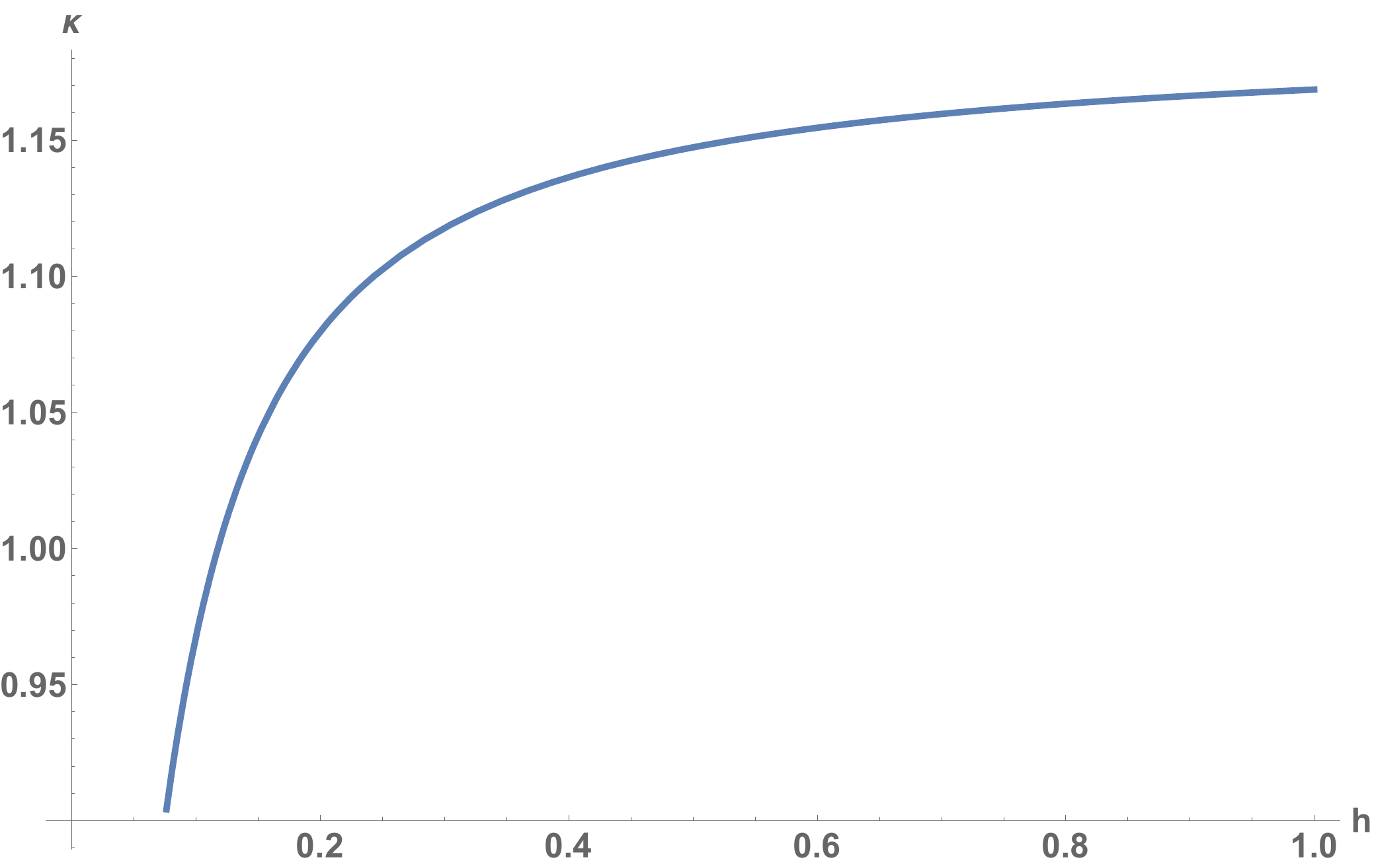}
\caption{The amount of $\mathcal{K}_{1}$ vs. $\tilde{h}$. By increasing $\tilde{h}$, $\mathcal{K}_{1}$ increases.} \end{figure}

\begin{figure}[H]
\centering
\includegraphics[scale=0.5]{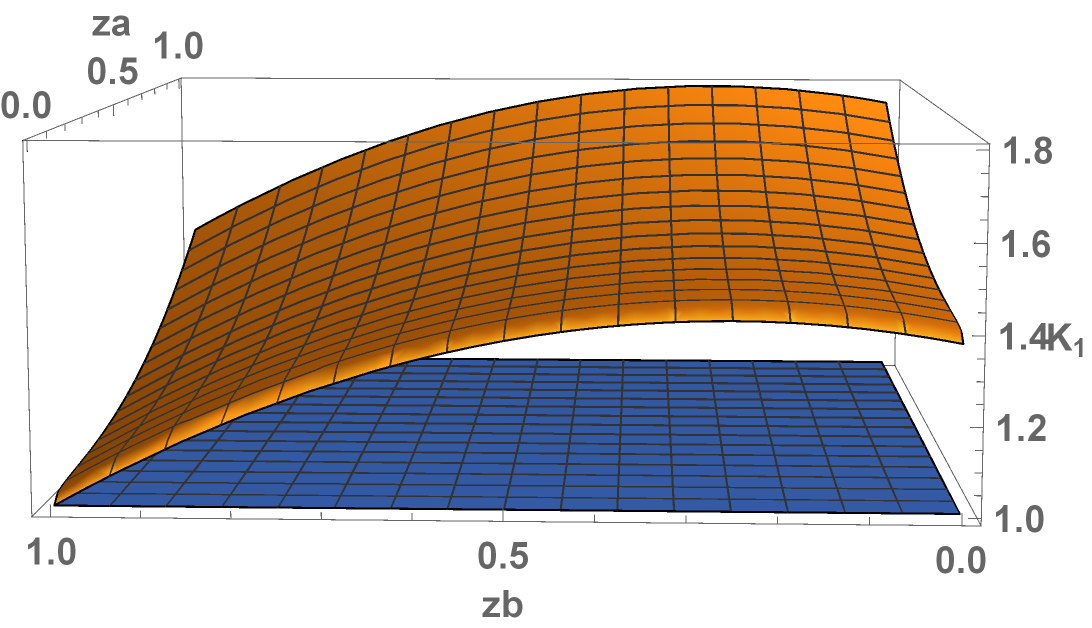}
\caption{The amount of $\mathcal{K}_{1}$ vs. $z_{a}$ and $z_{b}$. By considering $z_{c}=1$, the inequality will be violated for all other amounts of $z_{a}$ and $z_{b}$.} \label{fig6}
\end{figure}

\section*{Acknowledgement}
The authors are indebted to Prof. A.J. Leggett for his valuable remarks on an early draft of the paper.

\appendix
\section{Appendix A}
\renewcommand{\theequation}{A-\arabic{equation}} 
\setcounter{equation}{0}
\label{App A}
We calculate $\bra{0}\hat{U}_{vac}\ket{1}$ and $\bra{0}\hat{U}_{\alpha}\ket{0}$ here to show that they are approximately zero, even for asymmetric double-well potentials. First, for $\bra{}\hat{U}_{vac}\ket{1}$, we have:
\be
\label{A1}
\bra{0}\hat{U}_{vac}\ket{1}=-\dfrac{i}{\tilde{h}}\bra{0}\delta V(t)\ket{1}-\dfrac{1}{2\tilde{ h}}\bra{0}\mathfrak{g}\ket{1},
\ee
where $\mathfrak{g}$ is defined as:
\be
\label{A2}
\mathfrak{g}=-\frac{1}{2\tilde{h}}\sum_{\alpha}\int_{0}^{t}dt_{2}\int_{0}^{t_{2}}dt_{1}\hat{f}_{\alpha}(t_{2})e^{-i(t_{2}-t_{1})\omega_{\alpha}}\hat{f}_{\alpha}(t_{1}).\\
\ee
For the first term, one can show that it is equal to:
\begin{widetext}
\beq
\label{A3}
\dfrac{1}{2}\sum_{\alpha}\omega_{\alpha}^{2}\bra{0}f_{\alpha}^{2}\ket{1}&=&\dfrac{1}{2}\sum_{\alpha,m=0,1}\omega_{\alpha}^{2}\bra{0}f_{\alpha}\ket{m}\bra{m}f_{\alpha}\ket{1}\nonumber\\
&=&\dfrac{1}{2}\sum_{\alpha}\omega_{\alpha}^{2}(\bra{0}f_{\alpha}\ket{0}\bra{0}f_{\alpha}\ket{1}+\bra{0}f_{\alpha}\ket{1}\bra{1}f_{\alpha}\ket{1})\nonumber\\
&=&\dfrac{1}{2}\sum_{\alpha}\bar{\gamma}_{\alpha}^{2}\omega_{\alpha}^{2}(f_{00}.f_{01}+f_{01}.f_{11})=\dfrac{t}{\pi}(f_{00}.f_{01}+f_{01}.f_{11})\int_{0}^{\omega}\dfrac{d\omega}{\omega}J(\omega)\nonumber\\
&=&\dfrac{t}{\pi}\dfrac{f_{00}+f_{11}}{f_{01}\Omega_{10}}.\Gamma_{1},
 \eeq
 \end{widetext}
which is negligible, because $\Gamma_{1}t/\Omega_{10}\ll1$. The second term is also zero, because the following integrals have meaningful values, only when the terms in denominator are equal to zero (\textit{i.e.}, $\Omega_{10}+\omega_{\alpha}=0$), which is impossible since $\omega_{\alpha}>0$ and $\Omega_{10}>0$, so the entire term vanishes. To show this, we have 
\begin{widetext}
\beq
\label{A4}
\mathfrak{g}&=&-\dfrac{1}{2 \tilde{h}}\sum_{\alpha,m}\int_{0}^{t}dt_{2}\int_{0}^{t_{2}}dt_{1}\bra{0}f_{\alpha}(t_{2})\ket{m}\bra{m}f_{\alpha}(t_{1})\ket{1}e^{-i\omega_{\alpha}(t_{2}-t_{1})}\nonumber\\
&=&-\dfrac{1}{2\tilde{h}}\sum_{\alpha}\int_{0}^{t}dt_{2}\int_{0}^{t_{2}}dt_{1}e^{-i(\Omega_{10}-\omega_{\alpha})t_{1}}\bra{0}f_{\alpha}\ket{0}\bra{0}f_{\alpha}\ket{1}e^{-i\omega_{\alpha}t_{2}}\nonumber\\
&&+e^{i\omega_{\alpha}t_{1}}\bra{0}f_{\alpha}\ket{1}\bra{1}f_{\alpha}\ket{1}e^{-i(\Omega_{10}+\omega_{\alpha})t_{2}}\nonumber\\
&=&-\dfrac{1}{2\tilde{h}}\sum_{\alpha}\int_{0}^{t}dt_{2}(\dfrac{1}{-i(\Omega_{10}-\omega_{\alpha})}(e^{-i(\Omega_{10}-\omega_{\alpha})t_{2}}-1)\bra{0}f_{\alpha}\ket{0}\bra{0}f_{\alpha}\ket{1}e^{-i\omega_{\alpha}t_{2}}\nonumber\\
&&+\dfrac{1}{i\omega_{\alpha}}(e^{i\omega_{\alpha}t_{2}}-1)\bra{0}f_{\alpha}\ket{1}\bra{1}f_{\alpha}\ket{1}e^{-i(\Omega_{10}+\omega_{\alpha})t_{2}}\nonumber\\
&=&-\dfrac{1}{2\tilde{h}}\sum_{\alpha}\lbrace \bra{0}f_{\alpha}\ket{0}.\bra{0}f_{\alpha}\ket{1}\dfrac{1}{-i(\Omega_{10}+\omega_{\alpha})}\lbrace \dfrac{1}{-i\Omega_{10}}(e^{-i\Omega_{10}t}-1)+\dfrac{1}{-i\omega_{\alpha}}(e^{-i\omega_{\alpha}t}-1)\rbrace\nonumber\\
&&+\bra{0}f_{\alpha}\ket{1}\bra{1}f_{\alpha}\ket{1}\dfrac{1}{i\omega_{\alpha}}\lbrace\dfrac{1}{-i\Omega_{10}}(e^{-i\Omega_{10}t}-1)+\dfrac{1}{i(\Omega_{10}+\omega_{\alpha})}(e^{-i(\Omega_{10}+\omega_{\alpha})t}-1)\rbrace\rbrace\simeq 0.
\eeq 
\end{widetext}
For $\bra{0}\hat{U}_{\alpha}\ket{0}$, one can show that
\be
\label{A5}
\bra{0}\hat{U}_{\alpha}\ket{0}=\dfrac{i}{\sqrt{2\tilde{h}}}\int_{0}^{t}dt_{1}\bra{0}f_{\alpha}(t_{1})\ket{0}e^{i\omega_{\alpha}t_{1}}=i\dfrac{2\pi}{\sqrt{2\tilde{h}}}\bar{\gamma}_{\alpha}f_{mn}\mathcal{D}_{1}(\omega_{\alpha})e^{i\omega_{\alpha}t/2},
\ee
 where
 \be
 \label{A6}
  \mathcal{D}_{1}(\omega;t)=\dfrac{1}{2\pi}\dfrac{\sin(\omega t/2)}{\omega/2}.
  \ee
  Then
  \be
  \label{A7}
\vert\bra{0}\hat{U}_{\alpha}\ket{0}\vert^{2}=\dfrac{2t}{\tilde{h}}f_{00}^{2}\int_{0}^{\infty}d\omega J(\omega)\mathcal{D}_{2}(\omega),
\ee
where
\be
\label{A8}
\mathcal{D}_{2}(\omega;t)=\dfrac{1}{2\pi t}\lbrace\dfrac{\sin(\omega t/2)}{\omega/2}\rbrace^{2}.
\ee
We work in the principal time domain for which $\mathcal{D}_{2}(\omega,t)\sim\delta(\omega)$. So the relation (A-7) is equal to zero, since $J(0)=0$. So, the term $\bra{0}\hat{U}_{\alpha}\ket{0}$ could be neglected. The same situation holds for the element $\bra{1}\hat{U}_{\alpha}\ket{1}$ with relations similar to $\bra{0}\hat{U}_{\alpha}\ket{0}$.
\section{Appendix B}
\renewcommand{\theequation}{B-\arabic{equation}} 
\setcounter{equation}{0}
\label{App B}
 Here, we calculate the term ${P}_{L\vert R}$ as an instance. Other probabilities can be obtained in the same way. 
We need to calculate some terms at first and then put them in the main formula. To show this, we have:
\begin{eqnarray}
\label{B1}
P_{L\vert R}&=&a^{2}b^{2}\vert\langle 0\vert \hat{U}_{vac}\vert 0\rangle\vert^{2} +a^{\prime2}b^{\prime2}\vert\langle 1\vert \hat{U}_{vac}\vert 1\rangle\vert^{2}\nonumber\\
&&+ 2aa^{\prime} bb^{\prime}\Re\langle 0\vert \hat{U}_{vac}\vert 0\rangle^{\ast}\langle 1\vert \hat{U}_{vac}\vert 1\rangle+{a^{\prime}}^2b^{2}\vert\langle 1\vert \hat{U}_{\alpha}\vert 0\rangle\vert^{2},
\end{eqnarray}
The terms $\bra{0}\hat{U}_{vac}\ket{0}$, $\bra{1}\hat{U}_{vac}\ket{1}$, $\bra{1}\hat{U}_{\alpha}\ket{0}$ and $\bra{0}\hat{U}_{\alpha}\ket{1}$ are calculated in ~\cite{[16]}. So, we have
\be
\label{B2}
\vert\bra{0}\hat{U}_{vac}\ket{0}\vert^{2}=1,\quad\vert\bra{1}\hat{U}_{vac}\ket{1}\vert^{2}=e^{-\Gamma_{1}t}
\ee
\be
\label{B3}
\bra{m}\hat{U}_{\alpha}\ket{n}=i\dfrac{2\pi}{\sqrt{2\tilde{h}}}\bar{\gamma}_{\alpha}f_{mn}\mathcal{D}_{1}(\omega_{\alpha}+\Omega_{mn};t)e^{i(\Omega_{mn}+\omega_{\alpha})t/2},
\ee
where $\mathcal{D}_{1}(\omega;t)=\dfrac{1}{2\pi}\dfrac{\sin(\omega t/2)}{\omega/2}$. For $\vert\bra{1}\hat{U}_{\alpha}\ket{0}\vert^{2}$, we have:
\beq
\label{B4}
\vert\bra{1}\hat{U}_{\alpha}\ket{0}\vert^{2}&=&\dfrac{t\pi}{\tilde{h}}f_{10}^{2}\int_{0}^{\infty}d\omega J(\omega)\mathcal{D}_{2}(\omega-\Omega_{10})\nonumber\\
&=&\dfrac{t}{\tilde{h}}f_{10}^{2}J(\Omega_{10})=\Gamma_{1}t\simeq1-e^{-\Gamma_{1}t},
\eeq
where $\mathcal{D}_{2}(\omega-\Omega_{10})\sim\delta(\omega-\Omega_{10})$.

All the terms that produced by multiplying the terms containing $\hat{U}_{\alpha}$ are zero because there is $\dfrac{\Gamma_{1}}{\Omega_{10}}$ ratio in all of them. 

There is also one non-zero multiplying term as the following:
\be
\label{B5}
\Re\langle 0\vert \hat{U}_{vac}\vert 0\rangle^{\ast}\langle 1\vert \hat{U}_{vac}\vert 1\rangle=e^{-\Gamma_{1}t/2}\cos(\bar{\Omega}_{10}t).
\ee
Finally, we obtain:
\be
\label{B6}
P_{R\rightarrow L}=\cos^{2}\theta-\cos^{2}\theta\cos2\theta e^{-\Gamma_{1}t}-2\sin^{2}\theta\cos^{2}\theta \cos(\bar{\Omega}_{10}t) e^{-\Gamma_{1}t/2}.
\ee

\end{document}